\documentclass{aa}
\usepackage{txfonts}
\usepackage{stfloats}
\usepackage{xcolor}
\usepackage{float}
\usepackage{caption}
\usepackage{lscape}
\def\msun{$\mathrm{M}_{\odot}$}

\def\starlight{{\sc Starlight}}
\def\fado{{\sc FADO}}
\def\pegase{{\sc P\'EGASE}}

\def\D4000{$D_{4000}$}

\newfont{\nlx}{cmssdc10 scaled 900}
\newcommand{\brem}[1]{\textcolor{black}{\nlx #1}}

\def\rbulge{$\mathrm{R}_{\rm B}$}

\newcommand{\sbb}{mag/$\sq\arcsec$}
\newcommand{\dmb}{$<\!\!\!\delta\mu_{9{\rm G}}\!\!\!>$}

\def\RY{${\cal RY}$}

\def\mbstar{${\cal M}_{\star,\textrm{B}}$}
\def\mhole{${\cal M}_{\bullet}$}

\def\mstotal{${\cal M}_{\star,\textrm{T}}$}
\def\tsstar{$\Sigma_{\star}$}
\def\sstar{$\Sigma_{\star,\mathrm{B}}$}

\newfont{\hvss}{cmssdc10 scaled 1540}
\def\?{{\bf\color{red}?}}
\def\mbtmass{$\langle t_{\star,\textrm{B}} \rangle_{{\cal M}}$}
\def\mbzmass{$\langle Z_{\star,\textrm{B}} \rangle_{{\cal M}}$}

\begin{document} 

\title{On the genesis of spiral galaxies}
\subtitle{Classical and pseudo bulges as extremities of a continuous sequence}
   \author{
          Iris Breda\inst{\ref{IAA-CSIC}}         
          \and          
          Polychronis Papaderos\inst{\ref{IA-FCiencias},\ref{ULisbon}}
          }

\institute{Instituto de Astrof\'{i}sica de Andaluc\'{i}a - 
Glorieta de la Astronom\'{i}a, s/n, 18008 Granada, Spain \label{IAA-CSIC}
\and
Instituto de Astrofísica e Ciências do Espaço, Universidade de Lisboa, OAL, Tapada da Ajuda, PT1349-018 Lisboa, Portugal 
\label{IA-FCiencias}
         \and 
Departamento de Física, Faculdade de Ciências da Universidade de Lisboa, Edifício C8, Campo Grande, PT1749-016 Lisboa, Portugal \label{ULisbon}  
\\
             \email{breda@iaa.es}
}

   \date{Received ???; accepted ???}
   
\abstract
{A tantalizing enigma in extragalactic astronomy concerns the chronology and driving mechanisms of the build-up of late-type galaxies (LTGs). The standard scenario envisages two formation routes, with classical bulges (CBs) assembling first in a quick and violent quasi-monolithic episode followed by gradual disk assembly, and pseudo-bulge (PB) formation occurring gradually over the gigayear-long secular evolution of LTGs through gentle gas inflow from the disk and in situ star formation at its centre. An expectation from this antagonistic rationale is the segregation of present-day LTG bulges into two evolutionary distinct groups, which is in sharp contrast with recent observations.}
{The present study aims for a thorough investigation of the star formation history (SFH) of LTGs with its ultimate goal being to outline a coherent framework for the formation and evolution of spiral galaxies and their main stellar components.}
{Using population spectral synthesis models, we analyse the spatially resolved SFH of the bulge and the disk for 135 LTGs from the CALIFA survey that adequately cover the relevant range in LTG mass. Complementarily, characteristic physical properties of bulges and disks, such as mean colours, mass- and light-weighted stellar age and metallicity, and EW(H$\alpha$), were contrasted with predictions from evolutionary synthesis models. The latter adopted exponentially declining SFHs, with an $e$-folding time $\tau$ between 0.1 and 20 Gyr.}
{Analysis of the SFH of approximately half a million spaxels consistently reveals that the main physical and evolutionary properties of both bulges and disks show a continuous distribution across present-day total stellar mass \mstotal. The $e$-folding time $\tau$ in spiral galaxies with log(\mstotal) > 10 increases from the centre to the periphery, suggesting that these systems grow in an inside-out fashion. Quite importantly, the radial gradient of $\tau$ in gigayear over kiloparsec in an individual galaxy increases with increasing \mstotal, which is consistent with a high bulge-to-disk age contrast in high-mass spirals, while lower-mass LTGs display roughly the same $\tau$ throughout their entire radial extent, with intermediate mass galaxies in between. Predictions obtained through evolutionary synthesis are overall consistent with observed properties, such as mean colours, stellar ages, and H$\alpha$ equivalent widths. Finally, bulges and disks of higher mass galaxies exhibit shorter formation timescales as compared to their lower mass counterparts.}
{Collectively, the obtained results evince a coherent and unified picture for the formation and evolution of LTGs, in which PBs and CBs denote extremities of a continuous mass sequence. This analysis is consistent with the framework where bulges are assembled jointly with their parent disks by gradual inside-out growth, at a pace that is regulated by the depth of the galactic potential. In accordance with this is the fact that the revealed correlations are entirely devoid of a bimodality, which would be expected if CBs and PBs were to emerge from two distinct formation routes.}

\maketitle

\parskip = \baselineskip

\section{Introduction \label{intro}}

Despite numerous observational and theoretical studies over the past decades, the nature and evolutionary pathways of galaxy bulges remain enigmatic. In fact, a comprehensive review of the literature on this key subject results in conflicting evidence.

In the framework of the currently standard scenario \citep[][and references therein]{KorKen04}, bulges in late-type galaxies (LTGs) come in two evolutionary distinct flavours. These might be classified as classical bulges (CBs) or pseudo-bulges (PBs), which seemingly display pronounced differences in their spectrophotometric and kinematic properties. At prima facie, a typical CB is comparable to a dwarf elliptical galaxy, exhibiting a spheroidal morphology and an old, metal-enriched and dynamically hot stellar component. CBs typically display inwardly steeply increasing surface brightness profiles (SBPs) that can be adequately approximated by a high ($\gtrsim 2$) $\eta$ S\'ersic profile \citep{Sersic63}, and comply with the Kormendy scaling relations \citep{Kormendy77} established for regular elliptical galaxies \citep{FisDro10}. They typically host a super-massive black hole (SMBH) that tightly correlates with their stellar mass \mbstar, stellar velocity dispersion $\sigma_{\star}$, and optical luminosity, eventually manifesting itself as an active galactic nucleus (AGN) \citep{Ho2008,KormendyHo2013}.

In contrast, PBs are characterized by a young and less metal-enriched stellar component that can be well fit by a lower ($\la 2$) S\'ersic index, typically displaying a significant degree of rotation and ongoing star-forming (SF) activity \citep{DroFis07,FisDro10}. Despite them being less common, the presence of Seyfert activity in PBs has been observationally established \citep{Kor11,KormendyHo2013,Kotilainen16}. In addition, although several works report that PBs do not abide by the SMBH mass (\mhole) vs. $\sigma_{\star}$ relation typical for galaxies hosting a CB \citep[e.g.][]{Kor11}, a recent analysis \citep{Ben21} does not reveal different trends between \mhole\ and $\sigma_{\star}$ for a sample of disk galaxies encompassing both PBs and CBs, uncovering instead a single scaling relation describing their entire galaxy sample.


Conventionally, it is acknowledged that present-day bulges (and their parent galaxies) arise through two distinct formation pathways. The long-established scenario frames the formation of CBs within a two-stage galaxy formation process, with the bulge firstly emerging after a quick and early quasi-monolithic star-formation event \citep{Lar74} or mergers \citep{BBF92,Agu01,KelNus12} associated with intense starbursts \citep{Oka12} and followed by circumnuclear gas accretion from which the disk emerges. 
A distinct formation route is envisaged within the standard scenario for PBs, with preceding disk formation succeeded by the build-up of a bulge-like, central luminosity excess through quasi-continuous SF fed by mild gas inflow from the disk over the Gyr-long secular galaxy evolution \citep[e.g.,][]{CdJB96,Car01,KorKen04}. Additional processes shaping the properties of present-day PBs have been suggested, such as bar-driven gas inflow, inward stellar migration, minor mergers, CB SF-rejuvenation and/or dynamical re-arrangement of the disk's stellar populations \citep{SprHer05, Scannapieco10,Guedes13,Bird12,Roskar12,Grand14,Halle15,ThoDav06, Johnston12,Johnston14,Mor12,Mor16}.

In view of the two aforementioned bulge formation routes, one might anticipate a dichotomy in the spectrophotometric, chemodynamical and evolutionary properties of bulges in LTGs. However, observational evidence for such an explicit segregation remains elusive. In fact, there is solid evidence in the literature suggesting roughly simultaneous growth of CBs and their parent disks, as well as an equivalent number of studies suggesting that CBs are entirely formed early on, while the disk perpetuates its build-up process (this, however, implies gradual disk assembly around fully assembled bulges, which is not observationally established). For instance, in the work by \citet{MoHo06} the authors mention that most of the bulges in their sample cannot be properly reproduced by either collapse or merger models, falling between both, and that blue bulges might result from mergers or secular evolution. The authors additionally remark that the metallicities of all bulges are correlated with the same of their parent disks, perhaps hinting at a disk origin and/or the co-evolution of both stellar structures. In addition, a recent empirical study based on a combined analysis including spatially resolved spectral modelling of integral field spectroscopy (IFS) data and SDSS imaging data of a representative sample of the local LTG population extracted from the CALIFA survey \citep{Sanchez12-DR1,Sanchez16-DR3}, uncovered a continuity across three orders of magnitude in total galaxy mass in the main properties of bulges (namely, stellar mass and surface density, mean mass-weighted stellar age and metallicity, and mass-weighted stellar age gradients), which are ultimately dictated by the total stellar mass (\citeauthor{BP18} \citeyear{BP18}, hereafter BP18, and \citeauthor{Bre20} \citeyear{Bre20}).

The severe inconsistencies between the conclusions drawn by the various analyses are most likely resulting from differences in the selection of galaxy samples and from the different methodologies adopted for bulge-(bar)-disk decomposition \citep[see,][]{Cook20}, along with the uncertainties introduced by their inherent limitations. For instance, it is common practice to initiate the analysis by classifying LTGs according to the nature of their bulge, dividing them into CBs and PBs on the basis of their S\'ersic index $\eta$ ($\eta$ $\la$ 2 for PBs and $\eta$ $\ga$ 2 for CBs) or the Kormendy relation. However, since bulge classification on the basis of $\eta$ lacks robust physical foundation \citep{Gadotti12,FisDro10}, and $\eta$ itself greatly depends on the adopted image decomposition method and on the analysed photometric bands \citep{Men08,Pap22}, such pre-segregation of LTGs into CBs and PBs can introduce a bias that propagates in the further analysis. Furthermore, it is customary to assume that the stellar disk, which is antecedently subtracted from the galaxy SBP in order to isolate the emission of the bulge, conserves its exponential nature within the bulge. Such an assumption, which might not hold true for all spirals, bears serious implications that can affect all the subsequent analysis \citep[see,][]{Bre20b}. In addition, the strong degeneracy between the S\'ersic $\eta$ and the effective radius R$_{\rm eff}$ potentially prevents a robust determination of the S\'ersic parameters \citep[see e.g.,][]{iFIT}. Point spread function (PSF) convolution effects and the fact that different studies might be founded on different photometric passbands adds further ambiguities to the procedure. Regarding the chemical and evolutionary properties of bulges and disks, these can be estimated by using photometric colours, spectral energy distribution (SED) fitting or spectral modelling of optical spectra. However, strong degeneracies plague such methods (for instance, the infamous age-metallicity-extinction degeneracy, e.g., \citeauthor{Worthey94} \citeyear{Worthey94}, \citeauthor{dMeu14} \citeyear{dMeu14}, or \citeauthor{fado} \citeyear{fado} for an extensive discussion on this topic) rendering it difficult to robustly determine the galactic stellar properties. In addition, aperture biases afflict determinations appertaining to single-fibre spectroscopy, which, in principle, IFS data can overcome.

Finally, the standard hypothesis of monolithic vs. secular bulge formation, leading, respectively, to CBs and PBs, needs to be confronted with the more recent picture of bulge formation through inward migration and coalescence of massive ($\ga 10^{8-9}$ \msun) SF clumps from the disk \citep{Noguchi99,Bournaud07,Car07,Elm08,Mandelker14,Mandelker17}.
This process, which may be regarded as intermediate between the two aforementioned scenarios, may naturally account for various observational constraints that point to an interwoven bulge-disk co-evolution. In fact, bulge-disk co-evolution has been suggested multiple times, after uncovering significant affinity between the main properties of the two stellar structures \citep[e.g.,][]{GA01,Gad09,PelBal96,FerLor14,CdJB96,vD13}. For instance, BP18 found a strong correlation between bulge and disk chemical and evolutionary properties for all LTGs in their sample, firmly indicating that bulge and disk have evolved jointly. In addition, this study has found that the bulge-to-disk age and metallicity contrast increases with increasing galaxy mass, suggesting that whereas the centres of low mass LTGs are nearly identical to their parent disks, the centres of high-mass LTGs have formed earlier and quicker, having experienced a higher degree of chemical enrichment. The combined results are interpreted such that all LTGs start their evolutionary journey analogously to their low-mass counterparts, that is as a rotating, star-forming, exponential disk. Progressively, the bulge emerges from the central SF climax (as currently observed in low-mass LTGs), with its formation rate and/or timescale (slower or faster; earlier or later) being directly proportional to the total galaxy mass.

With these considerations in mind, the present study explores a representative sample of the local LTG population, constituted by 135 exemplars extracted from the CALIFA IFS survey \citep{Sanchez12-DR1,Sanchez16-DR3} that spans over three orders of magnitude in total stellar mass. It complements the previous analyses BP18 and \citet{Bre20}, aiming to dissect the two LTG's main stellar components and assess their origins and interdependence
through analysis of the star-formation history (SFH) of bulge and disk. Its prime motivation revolves around the fact that the average quantities discussed in BP18 might fail to correctly represent the mass assembly history (MAH) by, for instance, diluting extreme values (e.g., if the present-day stellar mass was assembled in two equivalently strong SF episodes separated by 10 Gyr, the resulting mean stellar age is unable to capture such a bimodal SFH, merely providing an estimate in between of the peak of the two SF episodes). Ergo, the previously obtained results and subsequent interpretations can only be crystallized after in-depth examination of the SFH, which contain discrete information of the MAH of the galaxy and of its stellar structures. 


Ultimately, part of the present article is dedicated into predicting several sub-galactic stellar properties through evolutionary synthesis techniques, namely \pegase\ 2 \citep{Pegase}, and contrast them with observations. Such predictions arise from models parametrized by observational constraints, such as the $e$-folding time inferred for the sample bulges and disks, by fitting exponentially declining $\tau$ models to the SFHs of the individual sub-galactic components. By inferring the MAH of bulges and disks through population spectral synthesis techniques\footnote{Some remarks on the complex relation between SFH and MAH are provided in Sect.\ref{disc}.}, and model and subsequently compare predictions given by an evolutionary synthesis tool with observations, the present exercise will provide the first empirical constraints on whether SFH parametrizations can adequately model the MAH of the sub-galactic stellar components under study.


%


The manuscript is organized as follows: Sect. \ref{meth} describes the employed methodology and obtained results, followed by Sect. \ref{disc} which is dedicated to the discussion, and Sect. \ref{conc} which summarizes the conclusions. The present work is contiguous to BP18 and \citet{Bre20}, to where the reader is referred to for a complete description of the sample selection and of the details of the spatially resolved spectral synthesis analysis. 

\section{Analysis outline and main results \label{meth}}

As aforementioned, to conduct this study it were used 135 local, non-interacting, face-on LTGs extracted from the CALIFA survey \citep{Sanchez12-DR1,Sanchez16-DR3}. The sample is representative of the local spiral galaxy population, considering that it encompasses LTGs with stellar masses distributed across three orders of magnitude, with present-day total stellar masses 8.89 $\le$ log(\mstotal) $\le$ 11.53 and present-day bulge masses 8.25 $\le$ log(\mbstar) $\le$ 11.24. According to the NASA/IPAC Extragalactic Database\footnote{http://ned.ipac.caltech.edu/} and after verifying by visual inspection, approximately half of the galaxy sample are barred. In respect to the main stellar properties of bulge and disk of the sample galaxies, there are no apparent fundamental differences between barred and respective non-barred counterparts.

Even though photometric decomposition is avoided due to the previously mentioned reasons, it is necessary to identify the spaxels pertaining to the bulge and the ones to the disk. Bulge delimitation resulting in an estimate for the isophotal bulge radius \rbulge\ was determined with the code {\sc iFit} \citep{iFIT}, being defined as the radius of the bulge's S\'ersic model measured at an extinction-corrected surface brightness of 24 $r$ \sbb. This resulted in the distribution 2\farcs5 $\le$ \rbulge\ $\le$ 11\farcs2, where only four galaxies (3\%) have a bulge diameter lower or equal than the CALIFA's point spread function (FWHM $\approx$ 2\farcs6). Simultaneously, spatially resolved (spaxel-by-spaxel) spectral modelling of the low-resolution (R $\sim$ 850) CALIFA IFS data in the V500 setup was carried on by means of the pipeline {\sc Porto3D} \citep{P13,G16} invoking the spectral fitting code \starlight\ \citep{Cid05} and a stellar library comprising simple stellar populations (SSPs) from \citet{BruCha03} for 38 ages between 1 Myr and 13 Gyr and four stellar metallicities (0.05, 0.2, 0.4 and 1.0 Z$_{\odot}$), referring to a Salpeter initial mass function and Padova 2000 tracks.

\begin{center}
\begin{figure}[t] 
\includegraphics[width=1\linewidth]{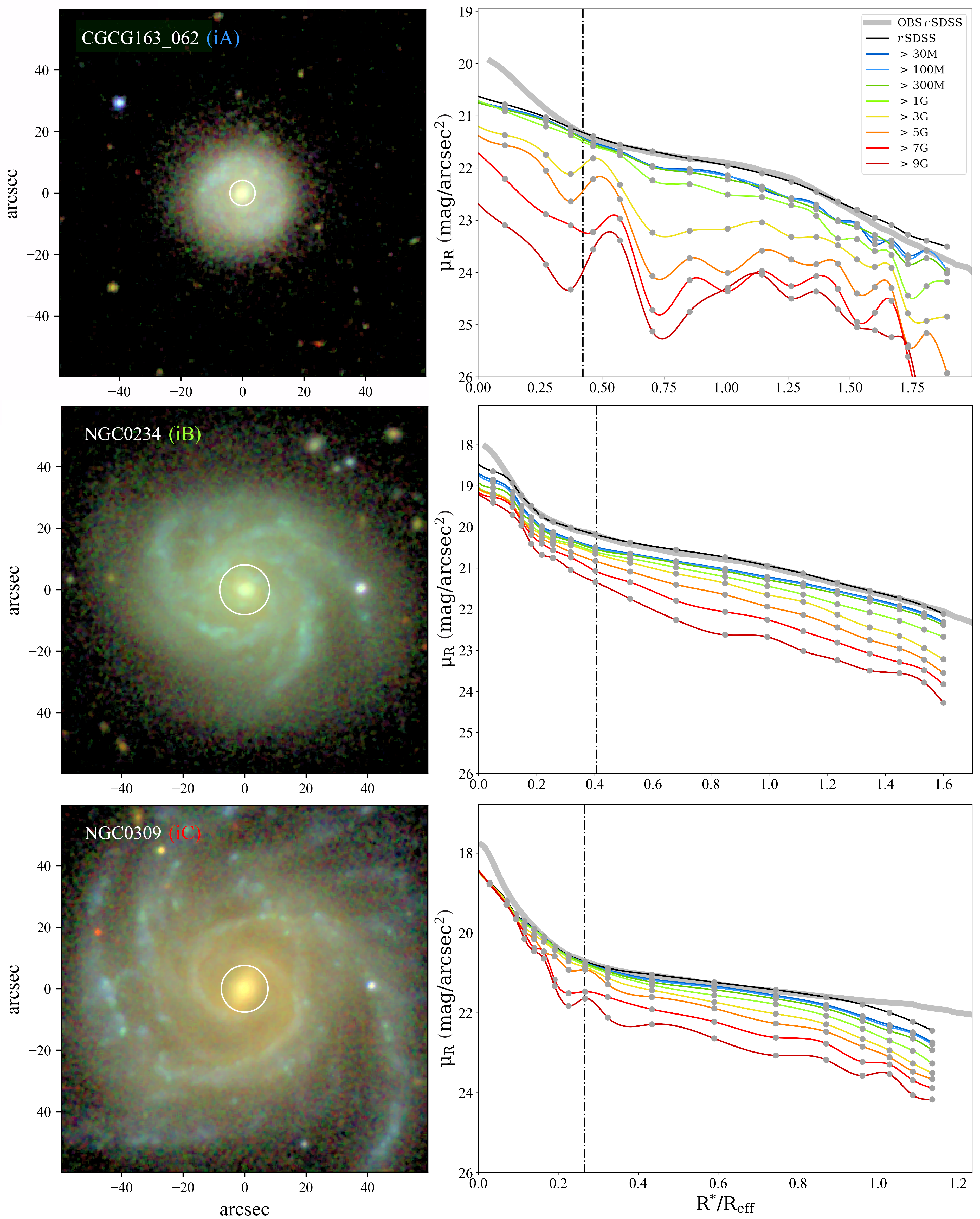}
\caption{SDSS true-colour images and respective SBPs for three exemplary LTGs, demonstrating the morphological differences of the three \dmb\ intervals tentatively defined in BP18. The synthetic SBPs were computed by convolving the observed IFS data with the SDSS $r$-band filter transmission curve after removal with \RY\ of stellar populations younger than 0.03, 0.1, 0.3, 1, 3, 5, 7, and 9 Gyr, being colour coded according to the age they represent. Shown in grey is the $r$-band SBP extracted from the SDSS photometric data. The vertical dashed line illustrates the estimated isophotal bulge radius \rbulge, depicted in the true colour image by a while circle.}\label{classes} 
\end{figure}
\end{center}
\vspace{-1.3cm}

Subsequently, the population vector (PV, i.e., the best-fitting combination of fractional contributions of individual SSPs to the galaxy light and mass) was post-processed with {\sc RemoveYoung}\footnote{The code and respective documentation is provided at \url{http://www.spectral-synthesis.org}} \citep[\RY;][]{GP16-RY}, a tool bound for striping the PV of the contribution from SSPs younger than an adjustable age cut-off $t_{\rm cut}$, additionally computing the synthetic spectrum, magnitudes in different filters (e.g., SDSS $u$, $g$, $r$, $i$, $z$) and stellar mass of the residual older stellar component. The spaxel-by-spaxel implementation of \RY\ for eight $t_{\rm cut}$ values (0.03, 0.1, 0.3, 1, 3, 5, 7, and 9 Gyr) resulted in surface brightness profiles ($\mu$) and stellar surface density (\tsstar) maps that were successively converted into 1D radial profiles, by means of the isophotal annuli (\brem{isan}) surface photometry technique \citep{P02}, and spline-interpolated to a finer radius.
The arithmetic average of the difference $\mu_{\rm 0\,Gyr}-\mu_{\rm 9\,Gyr}$, where $\mu_{\rm 0\,Gyr}$ and $\mu_{\rm 9\,Gyr}$ represent the individual synthetic $r$-band SBPs for a $t_{\rm cut}$ of 0 Gyr and 9 Gyr within \rbulge, respectively, was defined in BP18 as \dmb\ (mag). This is a distance- and formally extinction-independent estimate of the contribution of stellar populations of age $\geq$ 9~Gyr to the mean $r$-band surface brightness of the bulge. The LTGs are then classified into \dmb\ intervals, according to the approximately linear relation between \dmb\ and the bulge's mean mass-weighted stellar age \mbtmass\ (see panel $a)$ of Fig. 6 of BP18), resulting in three \dmb\ intervals:  interval~\brem{iA} (34 galaxies) with \dmb\ $\la$ --1.5 mag \&\ \mbtmass\ $\la$ 9 Gyr; interval~\brem{iB} (58 galaxies) defined by --1.5 mag $\la$ \dmb\ $\la$ --0.5 mag \&\ 9 $\la$ \mbtmass\ $\la$ 11 Gyr; interval~\brem{iC} (43 galaxies) falling in the locus \dmb\ $\ga$ --0.5 mag \&\ \mbtmass\ $\ga$ 11 Gyr. This is exemplified by Fig. \ref{classes}, which displays the true colour image of one exemplar from each \dmb\ interval, accompanied by their synthetic SBPs, computed through convolution of the observed IFS data with the SDSS $r$-band filter transmission curve (in black), and those after removal with \RY\ of stellar populations younger than 0.03, 0.1, 0.3, 1, 3, 5, 7, and 9 Gyr (from blue to red, respectively). Inspection of the PVs has demonstrated that \dmb\ tightly correlates with a typical mass fraction of stellar populations older than 9 Gyr. Specifically, a \dmb\ $\la$ --1.5 mag, as in the case of the bulge of ~\brem{iA} galaxies, corresponds to a mass fraction of about 20$\%$, whereas \brem{iB} display a typical value of $\sim$60$\%$, and ~\brem{iC} $\sim$85$\%$. A more detailed description of the photometric, spectral modelling and post-processing analyses is provided in BP18 which addresses how mean quantities, such as the bulge's mass-weighted stellar age and metallicity (\mbtmass\ and \mbzmass, respectively) and the bulge's stellar surface density \sstar, vary with the bulge stellar mass \mbstar. It additionally investigates how \mbstar\ relates to \mstotal. This exercise results in positive linear relations and utter absence of a dichotomy, strongly hinting at a unified formation scenario for LTGs. Secondly, it is worth noticing that attempting to segregate LTG bulges into PBs and CBs would result in a fair amount of bulges that would lie in between both groups, refraining from a sharp dichotomy to, at best, a smooth transition. Nonetheless, despite the consistency of the results obtained in BP18, mean quantities might be misleading. This stems from the fact that a different SFH can yield a similar mass- or light-weighted mean age estimate. In this regard, this work offers a closer examination of the SFHs of bulge and disk as obtained from PVs, aiming to gain further insights into the stellar mass growth of these two components.

\begin{center}
\begin{figure}[t] 
\includegraphics[width=1\linewidth]{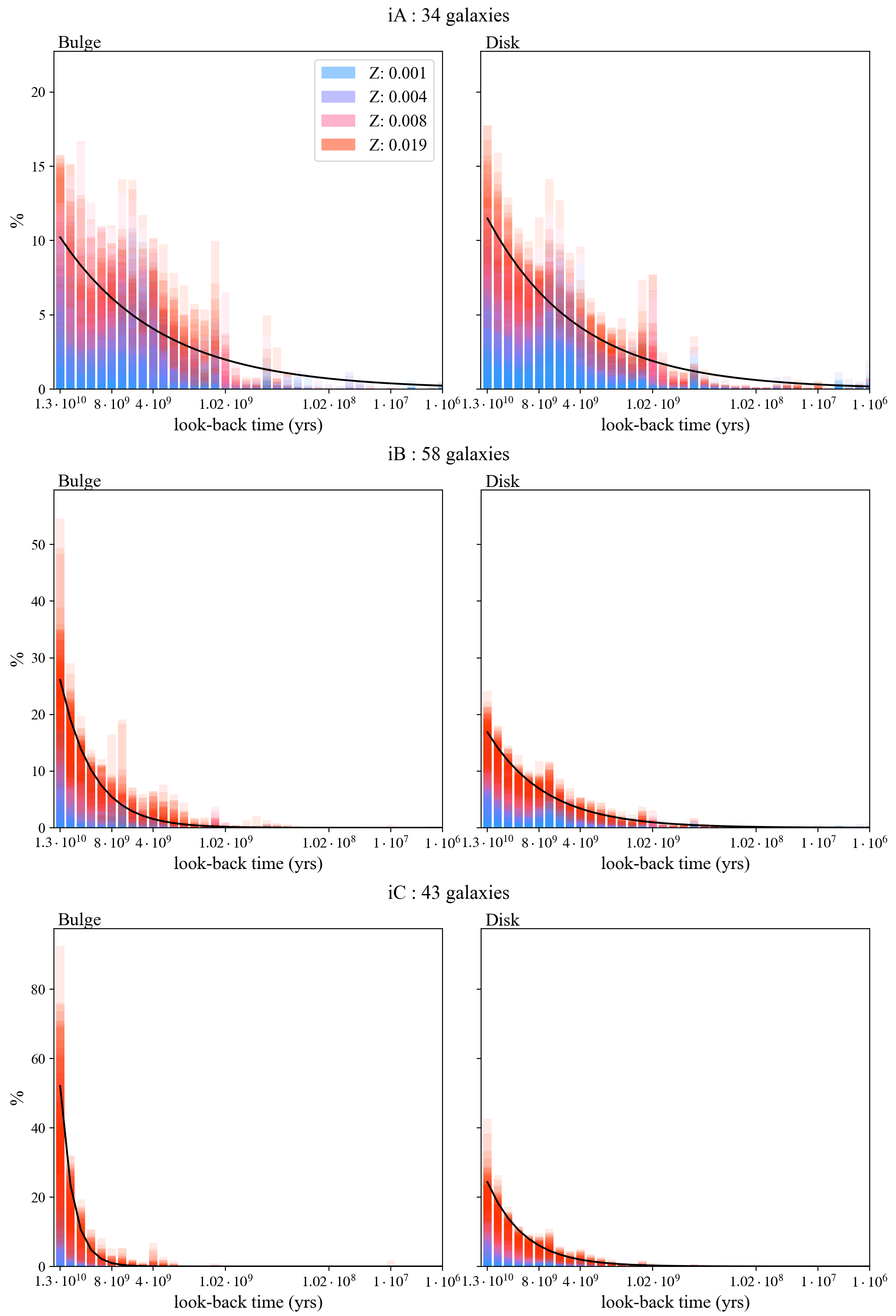}
\caption{
Series of histograms depicting the mass assembly through time (assembled ever-formed stellar mass vs. lookback time) of bulges (left-hand side) and disks (right-hand side) of the three \dmb\ intervals (\brem{iA}, \brem{iB} and \brem{iC}, from top to bottom). The histograms refer to the total stellar mass ever formed, not being corrected from the mass returned into the ISM through SNe and stellar winds. In addition, the histograms result from the combination of the mass assembly histories of all galaxies appertaining to each interval. It is colour coded according to the metallicity of each stellar population, as indicated by the legend.}\label{SFHs} 
\end{figure}
\end{center}     
\vspace{-1.3cm}

\begin{center}
\begin{figure}[t] 
\includegraphics[width=1\linewidth]{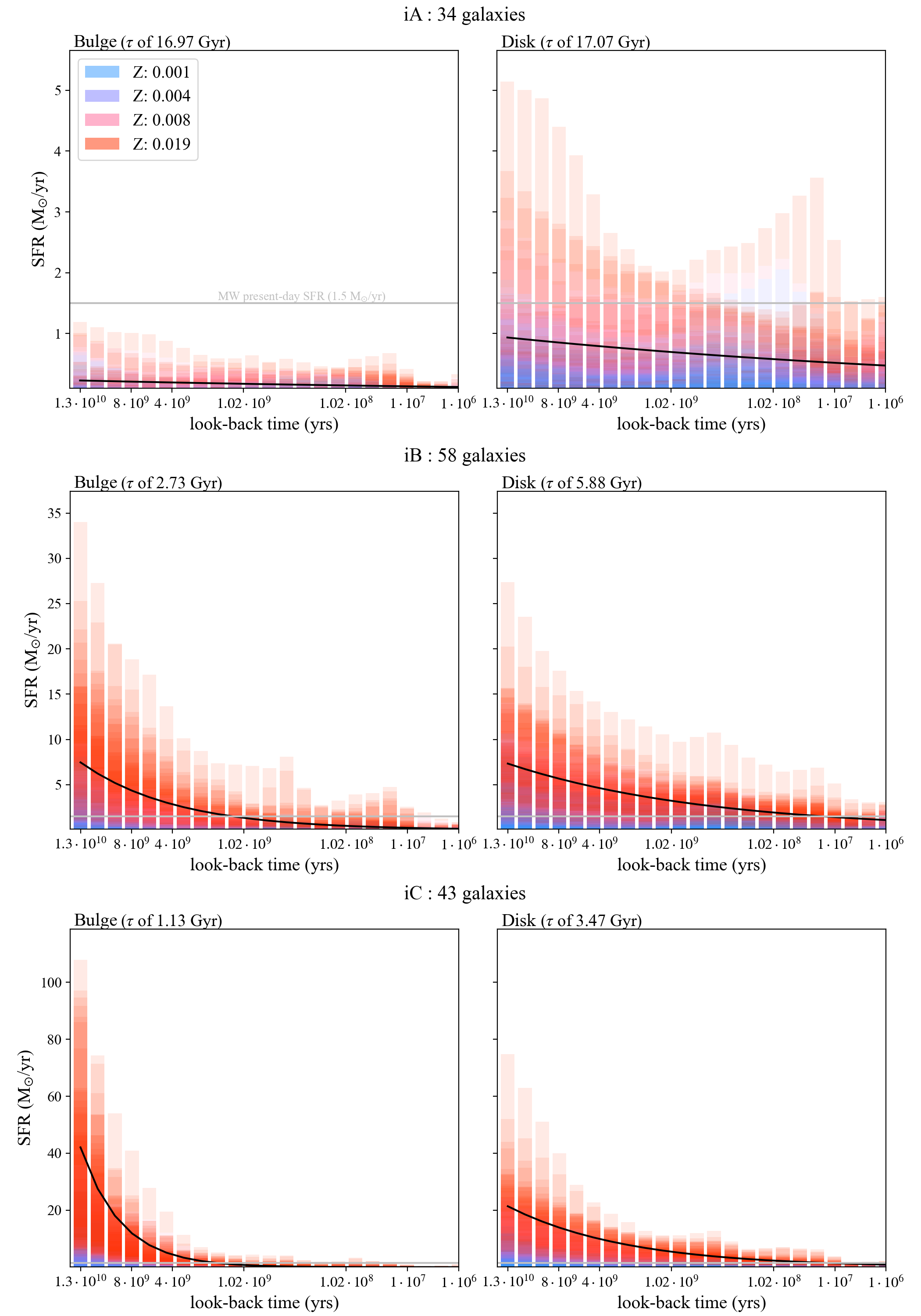}
\caption{
Distribution of the SFRs over look-back time averaged for each galaxy interval, in logarithmic scale. Exponential fits are over-plotted, highlighting the systematic decrease of the $e$-folding time of both structural components when transiting from the lower mass interval to the highest.}\label{SFRs} 
\end{figure}
\end{center}

Figure \ref{SFHs} displays the combined mass assembly histories of each \dmb\ interval, of both bulges and disks. As expected and already apparent from contiguous studies, it is revealed that whereas \brem{iC} bulges and their parent disks were assembled fast and early, \brem{iA} stellar components experienced (and still experience, considering their present-day SF activity) a more prolonged mass assembly, with \brem{iB} bulges and respective disks in between. Such behaviour is reaffirmed by Fig. \ref{SFRs} which displays the SFHs (i.e., the mean star-formation rates, SFRs, within individual time intervals across look-back time) averaged in each interval, for both bulges and disks. By assuming an exponentially declining SFR while performing exponential fits to the individual SFHs for each galactic component, the SFR $e$-folding timescales ($\tau$, i.e., the time interval in which the SFR decreases by a factor of \textit{e}) are obtained.

\vspace{-1.3cm}

\begin{center}
\begin{figure}
\includegraphics[width=1\linewidth]{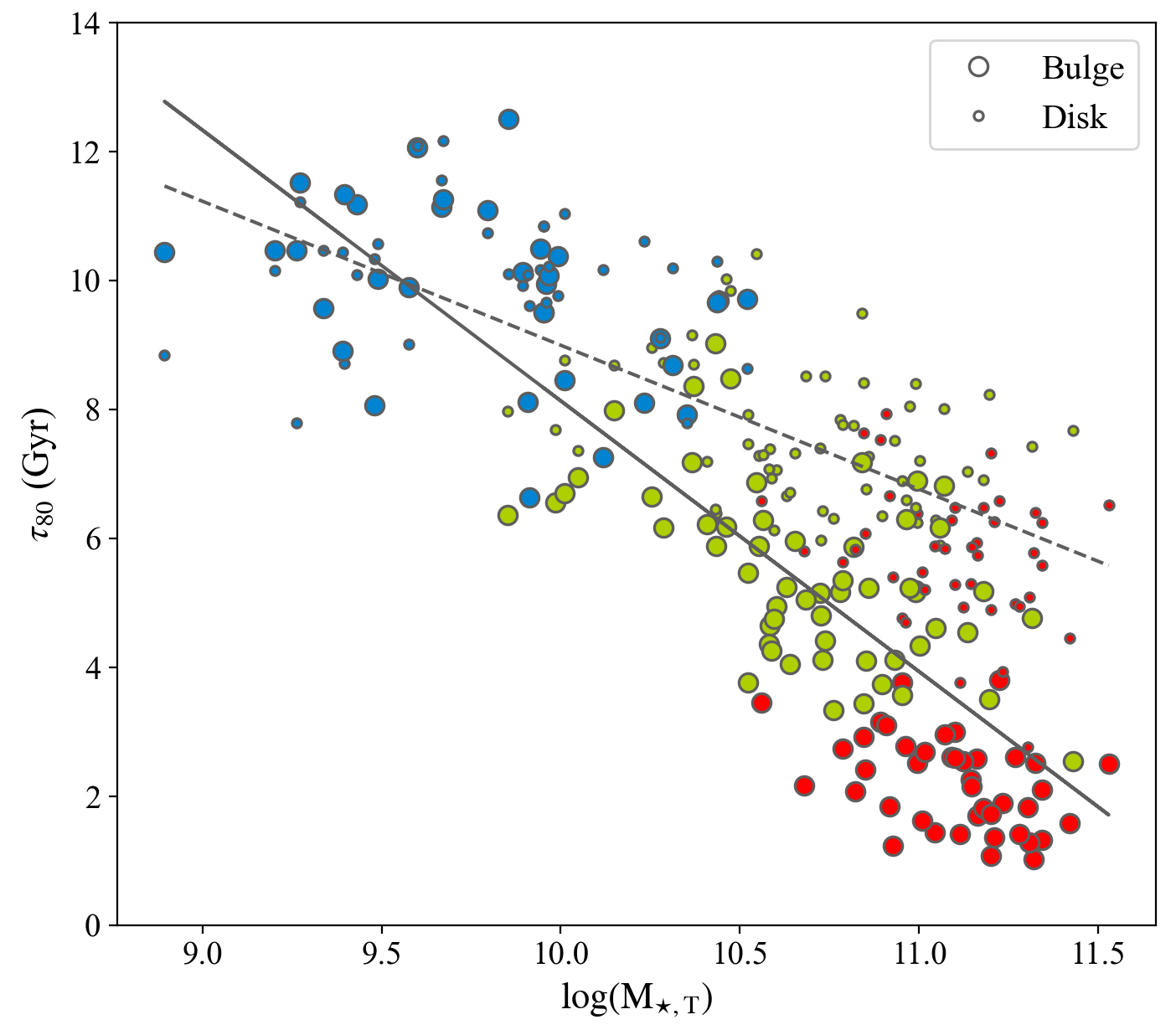}
\caption{
Time required to assemble 80\% of the ever-formed stellar mass of bulges and disks vs. the total present-day galaxy mass. Linear relations for both stellar components are over-plotted (solid lines for the bulge, dashed lines for the disk). The marks are colour coded according to their \dmb\ interval (\brem{iA} blue, \brem{iB} green, \brem{iC} red) and their size represents the measurements for bulge and disk (bigger and smaller circles, respectively).}\label{t80} 
\vspace{1cm}

\includegraphics[width=1\linewidth]{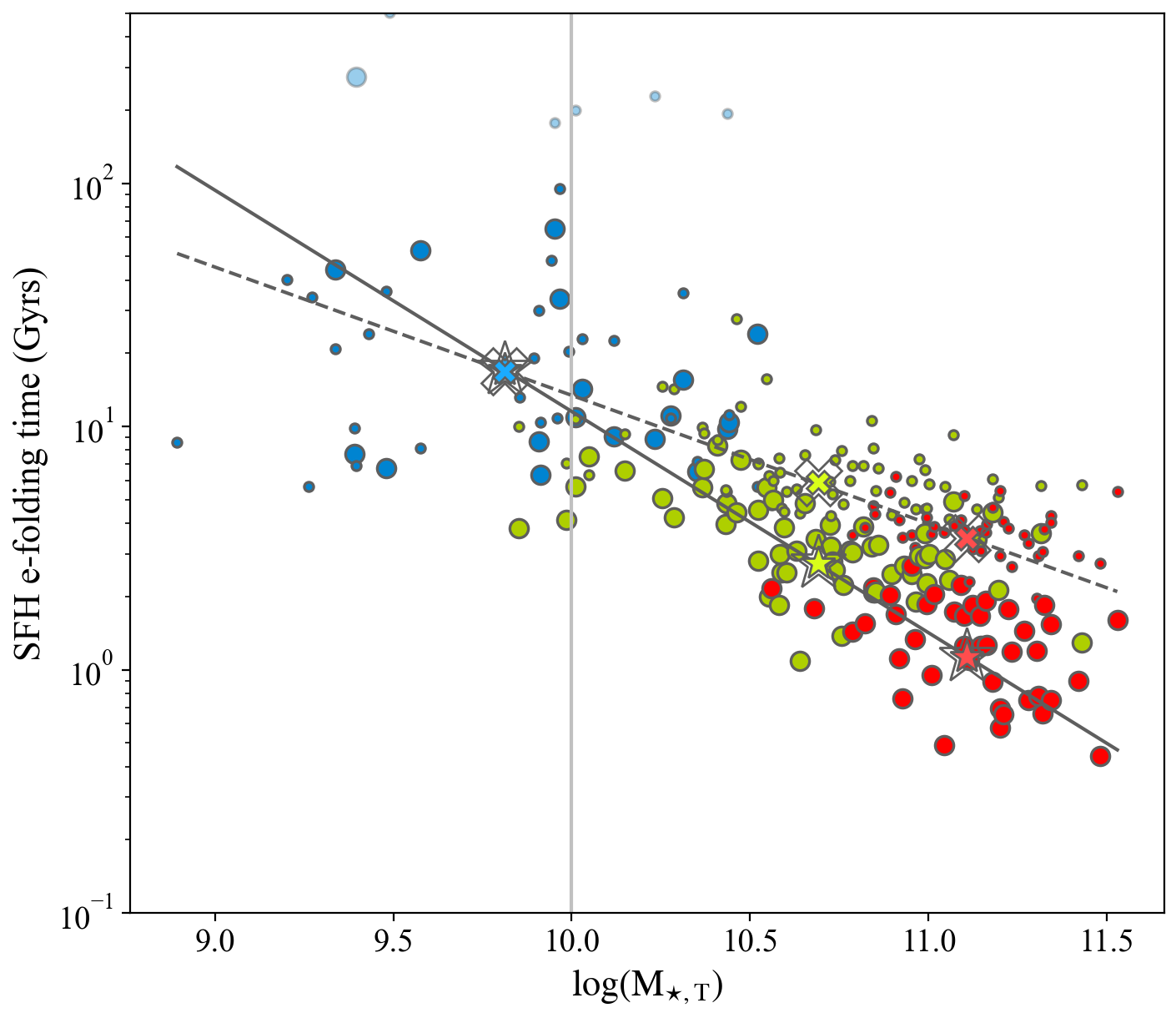}
\caption{
$e$-folding time of the SFH of both bulges and disks over the total present-day galaxy mass. Linear relations for both stellar components are over-plotted (solid lines for the bulge, dashed lines for the disk). The marks are colour coded according to their \dmb\ interval (\brem{iA} blue, \brem{iB} green, \brem{iC} red) and their size represents the measurements for bulge and disk (bigger and smaller circles, respectively). Average estimates for the $e$-folding times for both bulges and disks within each class are indicated by the coloured crosses and stars symbols, respectively. Transparent data points depict poor fits to the SFH distribution and are not considered in the linear fits.}\label{efold} 
\end{figure}
\end{center} 
\vspace{-1.3cm}

To better illustrate in what manner the total present-day stellar mass correlates with the pace at which the stellar mass assembly occurs, Fig. \ref{t80} displays the time required to assemble 80\% of the ever-formed stellar mass of the respective structural component, versus the total galaxy mass. Furthermore, Fig. \ref{efold} shows the $e$-folding time of the exponentially declining SFH over the total present-day galactic mass. To highlight how both distributions relate to each other, linear relations to bulge and disk are over-plotted by a solid and a dashed line, respectively. In accordance with the relations identified in contiguous works, this result uncovers linear correlations between the aforementioned quantities, yet again totally devoid of a segregation in two regions of the parameter space, as expected if bulges would emerge from the two antagonistic formation pathways envisaged within the standard scenario. In addition, it is apparent that whereas bulges and disks of \brem{iA} galaxies have assembled approximately at a similar pace, the disparity between the assembly timescale of both stellar structures gradually increases with increasing galaxy mass.

The linear fits modelling the dependence of the $e$-folding timescales for both stellar components on the total galaxy mass results in the following approximations, for bulge and disk, respectively:

\vspace{-0.5cm}
\begin{equation}
\mathrm{log}(\tau_{\mathrm{B}}) = -0.91 \, \mathrm{log}({\cal M}_{\star,\textrm{T}}) + 10.16,
\end{equation}

\vspace{-1cm}
\begin{equation}
\mathrm{log}(\tau_{\mathrm{D}}) = -0.53 \, \mathrm{log}({\cal M}_{\star,\textrm{T}}) + 6.40.
\end{equation}


As aforementioned, an additional element of this work consists in predicting several main sub-galactic properties, such as optical colours, mean stellar age and mean equivalent width of H$\alpha$ (EWH$\alpha$), by means of evolutionary synthesis modelling, and subsequently contrast the obtained estimates with observations. 

\begin{center}
\begin{figure}
\includegraphics[width=0.95\linewidth]{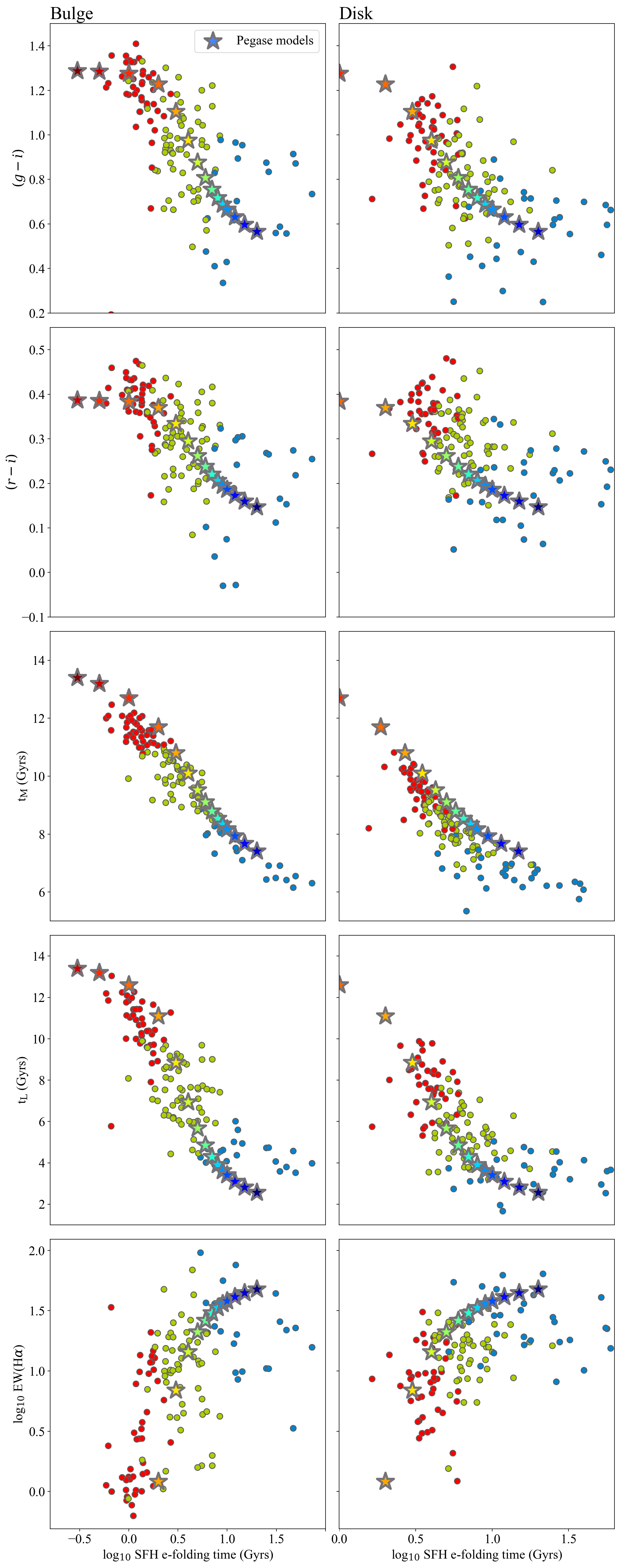}
\caption{Logarithm of the estimated $\tau$ for the sample bulges (left-hand side) and disks (right-hand side) vs. observed optical colours ($g$-$i$ in the first row and $r$-$i$ in second) after correction from intrinsic extinction effects, mean stellar age (mass-weighted in the third row and luminosity-weighted in the fourth), and EW(H$\alpha$) in the fifth row. Observations (defined by the circles) are contrasted with the model predictions (illustrated by the stars).}
\label{pred} 
\end{figure}
\end{center} 
\vspace{-1.3cm}

\begin{figure*}[b]
\begin{minipage}[t]{1\linewidth}
\includegraphics[width=\linewidth]{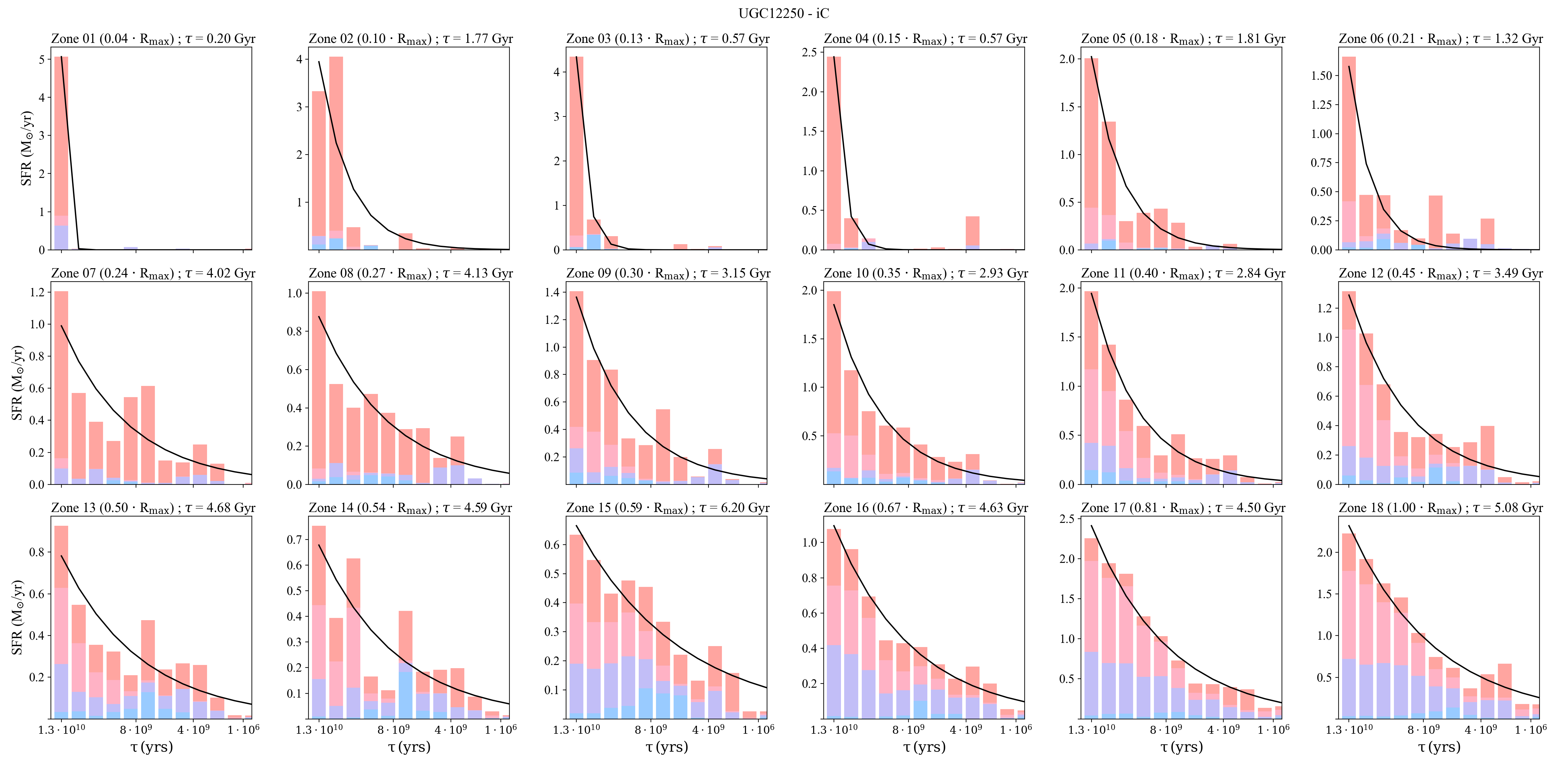}
\caption{SFHs of the individual zones the galaxy UGC 12250. The exponential fits are over plotted and the best-fitting $\tau$ is indicated on the upper part of each panel.}
\label{SFH_zones} 
\end{minipage}
\begin{minipage}[t]{0.49\linewidth}
\centering
\includegraphics[width=\linewidth]{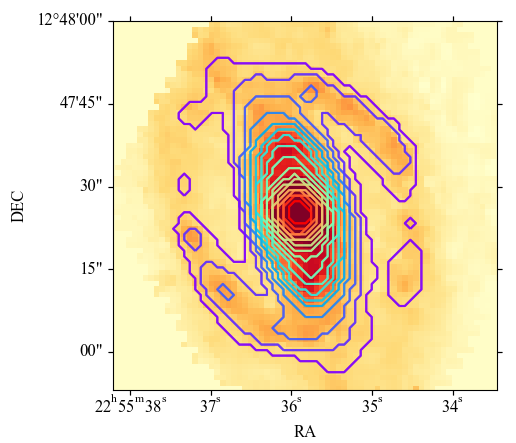}
\caption{Illustration of the logarithmically equidistant isophotal zones, i.e., \brem{isan}, for galaxy UGC12250 superimposed to the pseudo-continuum map. Each contour encompasses the spaxels pertaining to the respective \brem{isan}, tracing the galaxy morphology.}
\label{isan_ex}
\end{minipage}
\hspace{0.35cm}
\begin{minipage}[t]{0.49\linewidth}
\centering
\includegraphics[width=0.84\linewidth]{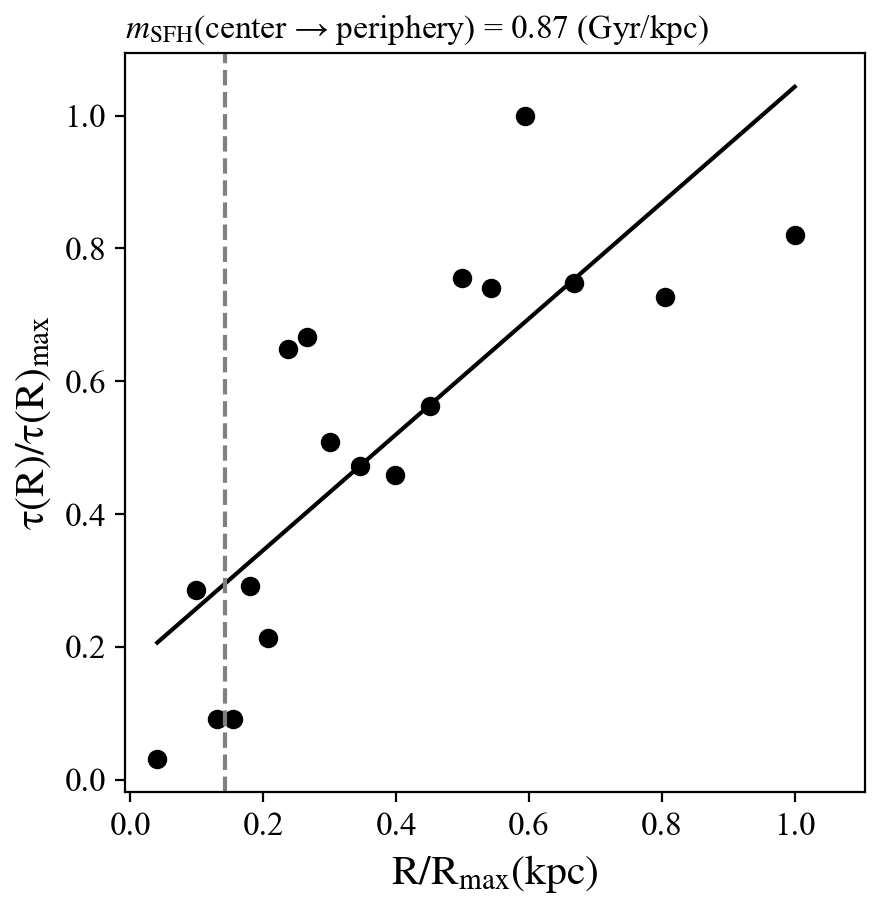}
\caption{Radial evolution of the best-fitting $\tau$ illustrating the nature of the SFHs of the individual zones (early collapse vs. continuous mass assembly) on the example of the galaxy UGC 12250. A linear fit was performed, which slope depicts how steeply the observed SFHs evolve, from the centre to the periphery of the galaxy. A grey vertical line indicates the isophotal radius of the bulge, comprising 3 \brem{isan}. Over the figure it is provided the value of the estimated slope.}
\label{SFH_slope}
\end{minipage}
\end{figure*}

By employing \pegase\ 2 \citep{Pegase} it was computed the synthetic SEDs for 90 ages between 0 and 13.7 Gyr. These predictions were obtained by assuming an exponentially declining SFH with varying $\tau$ (between 0.1 Gyr and 20 Gyr), a Salpeter IMF (between 0.1 and 100 M$_{\odot}$) and at fixed solar metallicity. 
By comparing the theoretical predictions with the observed stellar properties, the present exercise assists to evaluate the validity of our assumption, namely that the formation (i.e., MAH) of bulges and disks can be approximated by an exponentially declining SFH with a $\tau$ which is indirectly proportional to the total galaxy mass, as demonstrated in Fig.\ref{efold}.

Bulge and disk average optical colours were computed from the SDSS frames in the subsequent fashion: $i$) after isolating the two stellar structures in the images, apparent magnitudes of both stellar components were determined by performing statistics directly on the SDSS frames; $ii$) the former magnitudes were corrected for intrinsic extinction by adopting the best-fitting values obtained through spectral modelling of the integrated spectra of both stellar components \citep[cf.][for additional details on the construction of the integrated spectra and subsequent spectral synthesis]{Bre20b} and following the prescription by \citet{CCM89}, adopting a R$\rm{_V}$ = A$\rm{_V}$/E(B-V) of 3.1; $iii$) SDSS optical colours were obtained through the difference between the previously corrected apparent magnitudes.

Figure~\ref{pred} evidences a fair agreement between observations (circles) and \pegase\ predictions (stars) for both bulges and disks and for 
all physical properties under study. It is important to note, however, that some of the assessed properties, in particular colours and EW(H$\alpha$) values, display a considerable scatter. In addition, there is a slight displacement between predicted and observed colours of the oldest, more massive bulges of the sample, which might stem from underestimating the stellar metallicity of these bulges (in this study the SSP library is limited to 1 Z$_{\odot}$, yet it is possible that the most massive bulges display higher metal content). Finally, the panel referring to the mass-weighted mean stellar age of the disk reveals a minimal but systematic displacement between predictions and observations. In any case, and despite slight disagreements, the overall consistency between observations and predictions, along with the utter absence of a bimodal distribution strengths the notion for a unified scenario for the formation and evolution of late-type galaxies, where bulges form a continuous sequence of increasing (decreasing) age/formation-timescale with increasing (decreasing) galaxy mass, evolving in tandem with the surrounding disk.


Complementary, it was examined the SFHs within 18 logarithmically equidistant isophotal zones, which are defined from the reference image of the emission-line-free pseudo-continuum within 6390 -- 6490 \AA, faithfully reproducing the galaxy morphology, without preceding assumptions on galaxy structure (see BP18 for detailed description of the \brem{isan} technique). In this fashion, for each sample galaxy the radial evolution of the SFHs was obtained and analysed by appreciating how the best-fitting $\tau$ (normalized to its maximum value) varies from the centre to the periphery of the galaxy. Galaxy UGC12250 is provided as example -- Fig.~\ref{isan_ex} displays the reference image derived from the emission-line-free pseudo-continuum map superimposed by coloured contours which depict the various \brem{isan} encompassing the spaxels pertaining to the respective zones. The SFR of each \brem{isan} is assessed by integrating the SFRs of the individual spaxels contained within the respective zone, subsequently fitting an exponential function, as demonstrated by Fig.~\ref{SFH_zones}. Lastly and as displayed by Fig.~\ref{SFH_slope}, the radial evolution of the derived $\tau$'s is examined. Linear fits to the obtained best-fitting $\tau$ provide a suitable measure on how steeply the SFHs radially unfold (i.e., a highly negative slope indicates quick and early inside-out formation whereas $\leq$ 0 suggests a more continuous mass assembly throughout the galactic radial extent).

\begin{figure}[h]
\includegraphics[width=\linewidth]{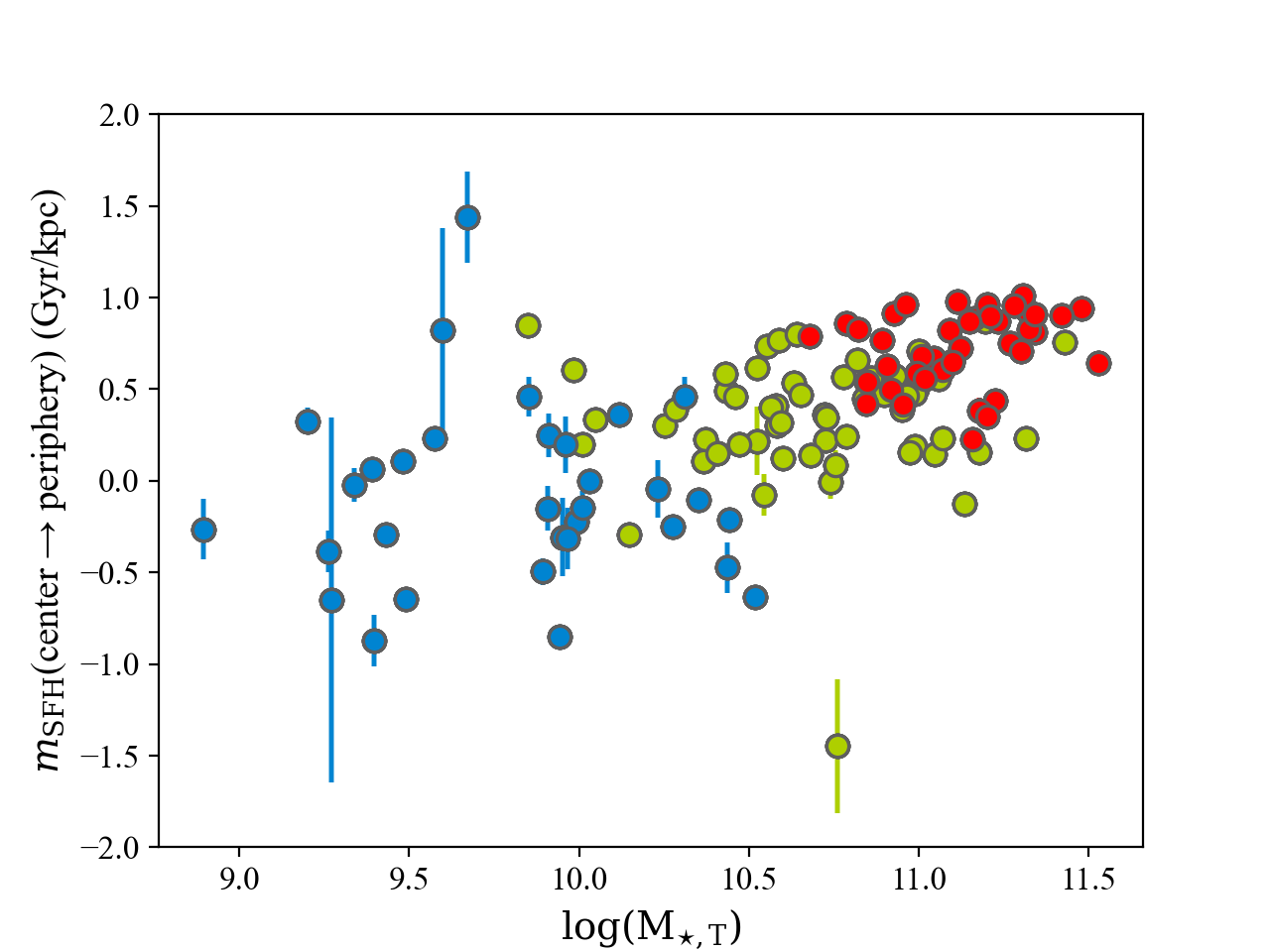}
\caption{Best-fitting slope of the $\tau$ radial distribution throughout the galaxy versus galaxy total mass. Analogously to the previous figures, data points are colour coded according to their \dmb\ interval (\brem{iA} blue, \brem{iB} green, \brem{iC} red).}
\label{slope_mass}
\end{figure}

Finally, the obtained slopes were correlated with the total stellar mass of each galaxy and the results can be appreciated in Fig.~\ref{slope_mass} were the $y$ axis represents how steeply the best-fitting $\tau$ for each SFH evolves, from the centre to the periphery of the galaxy, and the $x$ axis the total galaxy mass. From this figure it is evident that high-mass galaxies assembled their mass inside-out, rapidly and early, with the mass assembly in their centres occurring in an accelerated fashion as compared to its periphery. In contrast, the mass assembly in the central regions of low-mass spirals occurred roughly at the same pace as in their periphery, exhibiting a more continuous MAH throughout their radial extent. However, the fact that \brem{iA} bulges typically display positive mass-weighted stellar age gradients \citep{Bre20} also alludes to an inside-out formation scenario, considering that they are presently forming their centres more intensively as compared to the bulge periphery. 
Intermediate mass spirals fall exactly in between, as expected. Collectively, the obtained results delineate a coherent picture for the formation of disk galaxies, where the deepness of the potential well reflected by the total galaxy mass dictates the pace at which the galaxy will be assembled, from the centre towards the galactic outskirts.

\section{Discussion}\label{disc}


Preceding the following discussion it should be recalled that this work is based on spatially resolved spectral synthesis of 135 high-quality IFU CALIFA data cubes, of a representative sample of the local spiral galaxy population. By combining surface photometry, spectral modelling of IFS data and post-processing of population vectors with \RY, bulge and disk are individually investigated. Previous works based on the same data, such as BP18 and \citet{Bre20}, uncovered solid \textit{continuous} correlations between the total galaxy mass and several major physical properties of both disk and bulge, such as stellar mass, stellar surface density, mean stellar age and metallicity, and the slope of the bulge stellar age gradient. Herein, the individual SFHs of the sample bulges and disks are investigated and contrasted with predictions from evolutionary synthesis codes. Yet, it should be stressed that such task is non-trivial and that there are various uncertainties that must be considered, as reviewed in this subsection.


\subsection{SFH versus MAH}\label{SFHMAH}

Recent technological advancements have allowed for individual galaxies to be observed with unprecedented detail, leading to a substantial progress in our understanding of galaxy formation and evolution. Much of that knowledge derives from galactic archaeology, a technique that aims at studying the formation and evolution of galaxies by unravelling ages and chemical compositions of the different stellar populations. Herein, such is achieved by assessing the individual stellar populations through spatially resolved spectral synthesis, allowing the galaxy to be stripped of its young stellar populations, gradually revealing the morphology of older galactic structures. The underlying assumption, however, is that these old stellar structures roughly correspond to how these galaxies resembled in their past, i.e., that present-day's stellar populations are, on average, at nearly the same galactocentric radius as the one where they formed. Such assumption allows us to, for instance, conclude that galaxies are formed inside-out, in view of the gradual growth of the external galactic regions as observed through such techniques. An important remark worth contemplating is that, to assume that the distribution of the older stellar populations in the galaxy is roughly equivalent to the past galaxy structure is to postulate that the SFH and the MAH are comparable. Although those terms are being used in the literature interchangeably, these two processes might not be identical -- while the MAH indeed encapsulates the SFH, the opposite is not necessarily true. In fact, the only way that both histories are reciprocally analogous is if stellar populations preserve their initial orbits, i.e., the distance to the galactic centre, through their entire life span. With respect to galactic disks, various empirical analyses based on observations of individual Milky Way stars are revealing signatures of outward stellar migration, mostly from the inner to the outer disk \citep[e.g.,][]{Hay18,Fra18,Lian22}. Nevertheless, solid understanding of this effect and its details, such as what stellar populations/components are more/less prone to it, and, quite importantly, radial migration timescales, is still lacking.

\subsection{Technical limitations}

In the technical domain, it is essential to bear in mind potential limitations of the adopted techniques, and the extent of its ramifications. At first order, the S/N and spectral coverage of the available spectral data plays a role in the spectral synthesis. However, the satisfactory spectral resolution (850 < R < 1650) and spectral range (between 3700 and 7500 \AA) of the CALIFA data is expected to suffice for spectral synthesis tools to produce sensible results \citep{Car19}. In addition, it is reasonable to expect higher precision from a spaxel-by-spaxel analysis, as the one taken herein, in view that the obtained estimates result from typically hundreds individual spaxels that were fit in each isophotal annulus. Likewise, the adopted SSP library also impacts the obtained determinations, as shown in BP18 where we contrast the results based on two stellar libraries reaching different maximum stellar metallicities. Additional concerns regards the limited resolution for accurately resolving old stellar populations (the difference between the SEDs of stellar populations older than $\sim$5 Gy are minor, being increasingly difficult to properly disentangle populations on the upper red giant branch). Such a limitation prevents spectral synthesis from placing firm constraints on whether delayed SF scenarios (of 1 to 3 Gyr after the Big Bang) are, or not, in effect.

A challenging issue is the one of the age-metallicity degeneracy (the spectral signature of young/old stellar populations mimics the same of metal-poor/metal-rich, possibly leading to substantially different predictions of the SFH; e.g., \citealt{Worthey94,dMeu14}; see \citealt{fado} for a review). 
The requirement for consistency between hydrogen-line luminosities and EWs with those implied by the best-fitting SFH \citep[e.g.,][]{Gus01,fado} can alleviate this problem, although a limitation of this approach is the assumption of case B recombination. For instance, when observing a SF system through IFU data, crossing-over of ionizing photons from one spaxel to the adjacent ones is anticipated, which can affect spectral fitting solutions in various ways. Considering that, systematic use of \fado\ on spatially resolved SF spectra should be carried out with caution. Nonetheless, the effect from the age-metallicity-extinction degeneracy should mostly affect the spaxels which host significant SF activity \citep[see][for a comparison of the stellar properties as obtained with a purely stellar or self-consistent approach in normal SF galaxies, as observed by single-fibre SDSS spectral data]{Bre22}.


Finally, this study employs evolutionary synthesis to predict certain properties of the different galactic components such as bulge and disk (i.e., colours, mean stellar age and mean equivalent width of H$\alpha$). In fact, much of our understanding on galaxy evolution is based on the comparison between observed colours with predictions from evolutionary synthesis models. The latter use semi-empirical parametrizations of the SFH, according to which galactic disks are characterized by a nearly constant SFR (being equivalent to an exponentially delayed SFR with a long $e$-folding timescale; e.g., \citealt{Gal84}) whereas bulges are better described by an exponentially declining SFR with a $\tau$ that scales inversely with present-day stellar mass \citep{San86,GuiRoc87,Pog99,Gav02,Bic04}. It is however important to bear in mind that whereas observed colours for bulges, eventually after corrections for intrinsic extinction, broadly agree with predictions from exponential $\tau$ models \citep[or `delayed' $\tau$ models,][]{Pegase1}, these models are rather limited, being most likely unable to reproduce the sub-galactic mass assembly histories in its full complexity (see sub-Sect. \ref{SFHMAH}).

\subsection{Competing scenarios}


A scenario that is receiving increasing observational support \citep[e.g.,][]{Zan15} is that the formation of bulges in spiral galaxies is primarily driven by inward migration of SF clumps emerging out of violent disk instabilities, as proposed by several numerical studies \citep{Noguchi99,Bournaud07,Elm08}. Under this framework, early bulge assembly is driven by the primordial, gas-rich disk, through coalescence and subsequent inward migration of kiloparsec-sized gas clumps of $\sim$10$^8$ M$_{\odot}$. Indeed, clumpy disks with highly turbulent morphology are increasingly common at high-z ($\ge$ 2). Quite importantly, simulations suggest that bulges formed in this fashion closely mimic the properties and dynamics of classical bulges \citep{Elm08}, as formed through monolithic collapse independently of the disk, and showing early and sharp SFHs, concordant with our ~\brem{iC} bulges. In addition, these authors demonstrate that, in the event that the clump formation timescale in the disk is prolonged, a more extended bulge assembly history should be expected, being in fair agreement with the properties of ~\brem{iA} \& ~\brem{iB} bulges.

It is typically argued that the metallicity gradient imprinted on the stellar populations of the Galactic bulge, with the centre being populated by metal-rich, $\alpha$-enhanced stars, and the periphery (at higher latitudes, further from the galactic disk) by comparatively metal-poor stars with higher Fe abundances \citep[e.g.,][]{Gon11}, is solid proof of its dissipative origin \citep[e.g.,][]{Min95,Pip10}. Interestingly, however, several theoretical works demonstrate that the clump-migration scenario too leads to negative stellar-metallicity gradients and $\alpha$-enhancement, broadly similar to what is observed in the Galaxy \citep{Bou09,Gon11,Nat17}. In addition, thick disk stars seem to display a metal content comparable to the same of the old, metal-poor bulge (e.g., \citealt{Gon11}; BP18 additionally reports high affinity between the stellar populations of bulge and parent disk). The fact that bulge and host-disk display a similar chemical enrichment hints at a common formation route, as expected if both are formed from the same gas reservoir.
In contrast, scenarios where bulges were formed through dissipative collapse, independently of disks, do not offer any reasonable explanation for the consistent affinity observed between the two stellar components.

\section{Genesis of bulges in LTGs -- From a bimodal distribution to a continuous sequence}\label{conc}

The present study, which, apart from examining the SFHs estimated through spectral synthesis goes one step further by contrasting such estimates with predictions from evolutionary synthesis, reveals that the mass assembly rate and $e$-folding timescale of both bulges and disks scale inversely with total stellar mass.
Quite importantly, the uncovered correlations are completely devoid of a segregation in two regions of the parameter space, as should be expected if the bulges in the sample were emerging out of two antipodal formation routes (i.e., if bulges would genuinely come in two flavours, arising from two different formation scenarios, such segregation should be apparent in their scaling relations). To the contrary, all bulges display a continuous behaviour, alluding to a common formation scenario. Moreover, the disk properties are tightly related to the same of the bulge, strongly hinting at a nearly synchronous formation and joint evolution of both both stellar structures.

The complementary evolutionary synthesis analysis allows for a comparison between observed and predicted stellar properties (such as colours, mean stellar age and mean equivalent width of H$\alpha$), for both bulges and disks. Collectively, the various findings demonstrate that the evolution of both bulges and disks of LTGs is closely related to the total galactic stellar mass, analogously with the conventionally established downsizing scenario where lower mass galaxies are formed later and/or over larger periods of time, while high mass galaxies evolve earlier and faster. Likewise, bulges hosted by higher mass LTGs are assembled earlier and faster as compared with bulges in lower mass LTGs, being consistent with a sub-galactic downsizing scenario \citep[e.g.,][BP18]{NeiBosDek06,MouTan06}.

In essence, simulations involving bulge formation through disk instabilities seem to adequately explain the broad range of bulge properties that are here observed, as well as the continuous character of the uncovered correlations. Collectively, our findings favour a scenario where bulge and disk are concurrently formed, at a timescale that is inversely related to the total stellar mass. Through accumulation of matter from the disk in the galactic centre, high mass proto-disks with deep potential wells rapidly give rise to prominent bulges, whereas lower mass disks develop (fainter) PBs, overtime.

\begin{acknowledgements}
I.B. acknowledge financial support from the State Agency for Research of the Spanish MCIU through the “Center of Excellence Severo Ochoa” award to the Instituto de Astrofísica de Andalucía (SEV-2017-0709). P.P. was supported through FCT grants UID/FIS/04434/2019, UIDB/04434/2020, UIDP/04434/2020 and the project “Identifying the Earliest Supermassive Black Holes with ALMA (IdEaS with ALMA)” (PTDC/FIS-AST/29245/2017). The authors thank the anonymous referee for valuable comments and suggestions.
\end{acknowledgements}


\bibliographystyle{apa}



\end{document}